\documentclass[12pt]{article}

\global\arraycolsep=1pt

\usepackage{amsbsy,amssymb,latexsym,amsfonts, amsmath}
\usepackage{mathrsfs}
\usepackage{graphicx}

\numberwithin{equation}{section}
\newcommand{\bel}[1]{\begin{equation}\label{#1}}                     
\newcommand{\bal}[1]{\begin{eqnarray}\label{#1}}                     
\newcommand{\be}{\begin{equation}}
\newcommand{\ee}{\end{equation}}

\newcommand{\ex}{\mathrm{e}}
\newcommand{\de}{d}

\newcommand{\scr}{\scriptstyle}
\newcommand{\qq}{\qquad}

\renewcommand{\thefootnote}{\fnsymbol{footnote}}

\newcommand{\bea}{\begin{equation}}
\newcommand{\eea}{\end{equation}}

\begin{document}
%
%
\begin{titlepage}
\begin{flushright}
\normalsize
~~~~
OCU-PHYS 353\\
June, 2011\\
\end{flushright}

\vspace{15pt}

\begin{center}
{\LARGE  $A_{n}^{(1)}$ Affine  Quiver Matrix Model} \\
\end{center}

\vspace{23pt}

\begin{center}
{ H. Itoyama$^{a, b}$\footnote{e-mail: itoyama@sci.osaka-cu.ac.jp},
and
T. Oota$^b$\footnote{e-mail: toota@sci.osaka-cu.ac.jp}
}\\
%
\vspace{18pt}
%

$^a$ \it Department of Mathematics and Physics, Graduate School of Science\\
Osaka City University\\
\vspace{5pt}

$^b$ \it Osaka City University Advanced Mathematical Institute (OCAMI)

\vspace{5pt}

3-3-138, Sugimoto, Sumiyoshi-ku, Osaka, 558-8585, Japan \\

\end{center}
%
\vspace{20pt}
\begin{center}
Abstract\\
\end{center}
 We introduce $A_{n}^{(1)}$ $(n=1,2,\cdots)$ affine quiver matrix model
   by simply adopting the extended Cartan matrices as incidence matrices and
   study its  finite $N$ Schwinger-Dyson equations as well as their planar limit.
    In the case of $n=1$, we extend our analysis to derive the cubic planar loop equation for
 one-parameter family of models labelled by $\alpha$:
    $\alpha =1$ and $\alpha =2$ correspond to the non-affine $A_{2}$ case and the
     affine $A_{1}^{(1)}$ case respectively.
 In the case of $n=2$, we derive three sets of constraint equations for the resolvents
 which are quadratic, cubic and quartic respectively.


\vfill

\setcounter{footnote}{0}
\renewcommand{\thefootnote}{\arabic{footnote}}

\end{titlepage}

\renewcommand{\thefootnote}{\arabic{footnote}}
\setcounter{footnote}{0}

\section{Introduction}
\label{sec:intro}

  Schwinger-Dyson equation for  matrix models played an important role  in
 the development of 2d gravity and its extensions in nineties  and takes the form of infinite
 dimensional algebraic constraints \cite{FKNDVV,fromDavidtoIMatsu}.
 Among other things,  $A_{n}$ quiver (or conformal) matrix model was constructed such that it satisfies the $W_n$ constraints
automatically \cite{ITEP,Kostov}.
   The model in its $\beta$ deformation has contributed a great deal to
 the recent understanding of the connection \cite{AGTWill} between 2d conformal field theory and
 the instanton sum \cite{Nek}
  that derives the Seiberg-Witten curve \cite{SWandfollowups}.
  The understanding consists of the proof in some special cases \cite{proof},
  isomorphism of the curves in both sides \cite{DVIMO}
   and direct checks in the $q$-expansion 
   \cite{check,fromMMStoMMM}\footnote{For a more extensive list of references till now, see, for instance,
   \cite{IYone} as well as \cite{latest}.}.
  The case in which the incidence matrices take the generalized Cartan matrices of
  the affine Lie algebra $A_{n}^{(1)}$ appears to us an interesting and natural
  generalization and deserves study for its own sake.
  In this paper, we provide such model and
 study its finite $N$ Schwinger-Dyson equations and their planar limit.

   This paper is organized as follows. In the next section, we introduce one-parameter family of matrix models labelled by
       a  parameter $\alpha$ with two species
     of eigenvalues. The non-affine $A_{2}$ and  affine $A_{1}^{(1)}$ quiver matrix models correspond to
      the $\alpha =1$ and $\alpha =2$ cases respectively.
   In section three, we consider S-D equations of this ``$\alpha$" model. We consider the finite $N$
   S-D equations as well as their planar limit. We derive a cubic planar loop equation and the cubic curve
    associated with it.
     A drastic simplification is observed in the case where $\alpha =2$ and $W_{0} = -W_{1}$
      and the cubic symmetry of the curve is made manifest.
  In section four, we introduce an $A_{n}^{(1)}$ affine quiver matrix model. In section five, we derive
   the planar loop equations for the case of $n=2$. They take the form of quadratic, cubic and
    quartic constraints for the resolvents. In the appendix A, 
    we outline the derivation of the S-D equations in section three. The appendix B gives the detail of
    the derivation of the planar loop equation in section five.

While it is not unlikely that, with a proper engineering of the potential $W_{i}$ and
the choice of the contour \cite{DF,fromMMStoMMM}, the partition function
of the model may get identified with the Nekrasov function specified by a set of gauge theory data,  we are
 unable to find one so far.

\section{$A_{1}^{(1)}$ model and the $\alpha$ deformation}
 Following the punchlines in the introduction, let us consider the $\beta$-deformed matrix model
with the partition function
\be
\begin{split}
Z&:= \int \de^{N^{(0)}} \mu \int \de^{N^{(1)}} \nu
\prod_{1 \leq I<J \leq N^{(0)}} | \mu_I - \mu_J |^{2\beta}
\prod_{1 \leq I < J \leq N^{(1)}} | \nu_I - \nu_J|^{2 \beta}
\prod_{I=1}^{N^{(0)}} \prod_{J=1}^{N^{(1)}}
\frac{1}{|\mu_I - \nu_J|^{\alpha \beta}} \cr
& \qq \times
\exp\left( \frac{\sqrt{\beta}}{g_s} \sum_{I=1}^{N^{(0)}} W_0(\mu_I)
+ \frac{\sqrt{\beta}}{g_s}\sum_{J=1}^{N^{(1)}} W_1(\nu_J) \right).
\end{split}
\ee
 Here, we have left the range of the integrations unspecified
 except that  it is designed such that the integrand vanishes at the end points
 of the integrations. The second ``deformation" parameter $\alpha$
interpolates between the $\beta$-deformed matrix model of $A_1^{(1)}$ type ($\alpha=2$) and
that of $A_2$ type ($\alpha=1$).

For notational simplicity, let us introduce the ``effective" action
\be
\begin{split}
\ex^{-S_{\mathrm{eff}}}
&:= \prod_{1 \leq I<J \leq N^{(0)}} | \mu_I - \mu_J |^{2\beta}
\prod_{1 \leq I < J \leq N^{(1)}} | \nu_I - \nu_J|^{2\beta}
\prod_{I=1}^{N^{(0)}} \prod_{J=1}^{N^{(1)}}
\frac{1}{|\mu_I - \nu_J|^{\alpha\beta}} \cr
& \qq \times
\exp\left( \frac{\sqrt{\beta}}{g_s} \sum_{I=1}^{N^{(0)}} W_0(\mu_I)
+ \frac{\sqrt{\beta}}{g_s}\sum_{J=1}^{N^{(1)}} W_1(\nu_J) \right).
\end{split}
\ee

\section{S-D equation of the ``$\alpha$ model" and the planar limit} 
 Let us begin with the Virasoro constraints:
\be
0 = \int \de^{N^{(0)}} \mu \int \de^{N^{(1)}} \nu
\sum_{I=1}^{N^{(0)}} \frac{\partial}{\partial \mu_I}
\left( \frac{1}{z - \mu_I} \ex^{-S_{\mathrm{eff}}} \right),
\ee
we have
\bel{LEQ0}
\begin{split}
& \left\langle\!\!\!\left\langle
\sum_{I=1}^{N^{(0)}} \frac{1}{(z-\mu_I)^2}
\right\rangle\!\!\!\right\rangle
+ 2 \beta \left\langle\!\!\!\left\langle
\sum_{I=1}^{N^{(0)}} \sum_{\stackrel{\scr J=1}{( J \neq I)}}^{N^{(0)}}
\frac{1}{z-\mu_I} \frac{1}{\mu_I - \mu_J} 
\right\rangle\!\!\!\right\rangle \cr
& - \alpha \beta 
\left\langle\!\!\!\left\langle
\sum_{I=1}^{N^{(0)}} \sum_{J=1}^{N^{(1)}}
\frac{1}{z-\mu_I} \frac{1}{\mu_I - \nu_J} 
\right\rangle\!\!\!\right\rangle
+ \frac{\sqrt{\beta}}{g_s} 
\left\langle\!\!\!\left\langle
\sum_{I=1}^{N^{(0)}} \frac{W_0'(\mu_I)}{z-\mu_I}
\right\rangle\!\!\!\right\rangle = 0.
\end{split}
\ee
Similarly, from 
\be
0 = \int \de^{N^{(0)}} \mu \int \de^{N^{(1)}} \nu
\sum_{J=1}^{N^{(1)}} \frac{\partial}{\partial \nu_J}
\left( \frac{1}{z - \nu_J} \ex^{-S_{\mathrm{eff}}} \right),
\ee
we have
\bel{LEQ1}
\begin{split}
& \left\langle\!\!\!\left\langle
\sum_{J=1}^{N^{(1)}} \frac{1}{(z-\nu_J)^2}
\right\rangle\!\!\!\right\rangle
+ 2 \beta \left\langle\!\!\!\left\langle
\sum_{I=1}^{N^{(1)}} \sum_{\stackrel{\scr J=1}{( J \neq I)}}^{N^{(1)}}
\frac{1}{z-\nu_I} \frac{1}{\nu_I - \nu_J} 
\right\rangle\!\!\!\right\rangle \cr
& - \alpha \beta 
\left\langle\!\!\!\left\langle
\sum_{I=1}^{N^{(0)}} \sum_{J=1}^{N^{(1)}}
\frac{1}{z-\nu_J} \frac{1}{\nu_J - \mu_I} 
\right\rangle\!\!\!\right\rangle
+ \frac{\sqrt{\beta}}{g_s} 
\left\langle\!\!\!\left\langle
\sum_{J=1}^{N^{(1)}} \frac{W_1'(\nu_J)}{z-\nu_J}
\right\rangle\!\!\!\right\rangle = 0.
\end{split}
\ee
Let
\be
\widehat{\omega}_0(z):= \sqrt{\beta} g_s \sum_{I=1}^{N^{(0)}}
\frac{1}{z-\mu_I}, \qq
\widehat{\omega}_1(z):= \sqrt{\beta} g_s \sum_{J=1}^{N^{(1)}}
\frac{1}{z-\nu_J}.
\ee
Adding $g_s^2 \cdot$\eqref{LEQ0} and $g_s^2 \cdot$\eqref{LEQ1}, we have
\be
\begin{split}
& \left\langle\!\! \left\langle \bigl( \widehat{\omega}_0(z) \bigr)^2
\right\rangle\!\! \right\rangle
+ \left\langle\!\! \left\langle \bigl( \widehat{\omega}_1(z) \bigr)^2
\right\rangle\!\! \right\rangle
- \alpha \left\langle\! \left\langle 
\bigl. \widehat{\omega}_0(z) \widehat{\omega}_1(z) \bigr.
\right\rangle\! \right\rangle + 
\epsilon 
\left\langle\!\left\langle \bigl. \widehat{\omega}_0'(z) 
\bigr. \right\rangle \! \right\rangle
+ \epsilon \left\langle \!\left\langle 
\widehat{\omega}_1'(z) \right\rangle \! \right\rangle \cr
& + W_0'(z) \left\langle\! \left\langle 
\bigl. \widehat{\omega}_0(z) \bigr.
\right\rangle\! \right\rangle
+ W_1'(z) \left\langle\! \left\langle 
\bigl. \widehat{\omega}_1(z) \bigr.
\right\rangle\! \right\rangle 
 -  \left\langle\!\!\left\langle 
\bigl.   \hat{f}_0(z)     \bigr.
\right\rangle\!\!\right\rangle
 -  \left\langle\!\! \left\langle 
\bigl.  \hat{f}_1(z) \bigr.
\right\rangle\!\! \right\rangle   = 0
\end{split}
\ee
where 
\be
\epsilon:= \left( \sqrt{\beta} - \frac{1}{\sqrt{\beta}} \right) g_s,
\ee
\be
\hat{f}_0(z):= \sqrt{\beta} g_s 
\sum_{I=1}^{N^{(0)}} \frac{W_0'(z) - W_0'(\mu_I)}{z - \mu_I},
\ee
\be
\hat{f}_1(z):= \sqrt{\beta} g_s 
\sum_{J=1}^{N^{(1)}} \frac{W_1'(z) - W_1'(\nu_J)}{z - \nu_J}.
\ee

  To take the planar limit $g_s \rightarrow 0$, let
\be
\omega_{0,1}(z) := 
\lim \left\langle\!\left\langle  \widehat{\omega}_{0,1}(z) \right\rangle\!\right\rangle,
 \;\;\;
 f_{0,1}(z) := 
\lim \left\langle\!\!\left\langle  \hat{f}_{0,1}(z) \right\rangle\!\!\right\rangle,
\ee
We obtain 
\be
 \omega_{0}(z)^2+ \omega_{1}(z)^2 - \alpha \omega_{0}(z) \omega_{1}(z) 
 + W_0'(z) \omega_{0}(z) + W_1'(z) \omega_1(z)  - f_0(z) - f_1(z) = 0.
\ee

 Let us turn to the higher order constraints. In particular,  let us consider
 \bel{higherSD1}
0 = \int \de^{N^{(0)}} \mu \int \de^{N^{(1)}} \nu
\sum_{I=1}^{N^{(0)}} \frac{\partial}{\partial \mu_I}
\left( \frac{1}{z - \mu_I}
\sum_{K=1}^{N^{(1)}}  \frac{1}{\mu_I - \nu_K}
 \ex^{-S_{\mathrm{eff}}} \right),
\ee
 as well as
 \bel{higherSD2}
0 = \int \de^{N^{(0)}} \mu \int \de^{N^{(1)}} \nu
\sum_{K=1}^{N^{(1)}} \frac{\partial}{\partial \nu_K}
\left( \frac{1}{z - \nu_K} 
    \sum_{I=1}^{N^{(0)}}  \frac{1}{\nu_K - \mu_I}
\ex^{-S_{\mathrm{eff}}} \right).
\ee
 Taking the difference of these two equations  and carrying out some algebra which is outlined
 in the appendix A, we obtain
 
 \bel{ABCDanswer}
 \left\langle\!\!\!\left\langle
   \frac{2(1-\alpha)}{\alpha} A  - \alpha \beta B - \frac{(1- \alpha \beta)}{2} C  + D 
   \right\rangle\!\!\!\right\rangle   = 0,
\ee
where
\bel{ABCDdef}
\begin{split}
 A& : = 
 -\frac{1}{2} \left(\sum_{I \neq J} \frac{1}{z- \mu_I} \frac{1}{z- \mu_J }\right)^{\prime} 
 + (2\beta -1) \sum_{I \neq J}\frac{1}{z- \mu_I}\frac{1}{(\mu_I- \mu_J)^2}   \cr
 &+ \frac{2\beta}{3} \sum_{I \neq J \neq K \neq I} \frac{1}{(z- \mu_I)(z- \mu_J)(z- \mu_K)} \cr
 &+ \frac{\sqrt{\beta}}{g_s}\frac{W_0^{\prime}(z)}{2}  
   \left(\sum_{I \neq J} \frac{1}{z- \mu_I} \frac{1}{z- \mu_J }\right)
 - \frac{\sqrt{\beta}}{g_s} \left(\sum_{I \neq J} 
 \frac{W_0^{\prime}(z) - W_0^{\prime}(\mu_I)}{z - \mu_I} 
 \frac{1}{\mu_I - \mu_J}\right)   \cr
 & - \left( \mu_I \leftrightarrow  \nu_I,  W_0 \leftrightarrow W_1 \right), \cr
 B& : = \sum_{I, J, K} \frac{1}{(z- \mu_I)(z- \nu_J)(z- \nu_K)}
       -  \sum_{I, J, K} \frac{1}{(z- \mu_I)(z- \mu_J)(z- \nu_K)}, \cr
C & : = \sum_{I,K} \frac{1}{(z- \mu_I)(z- \nu_K)^2}  -  \sum_{I,K} \frac{1}{(z- \mu_I)^2(z- \nu_K)} \cr
 &  + \frac{1}{\alpha \beta}
 \left\{ \sum_I \frac{1}{(z-\mu_I)^2} + 
 2\beta \sum_{I \neq K} \frac{1}{z- \mu_I} \frac{1}{\mu_I -\mu_K}
 + \frac{\sqrt{\beta}}{g_s} \sum_I \frac{W_0^{\prime}(\mu_I)}{z- \mu_I} \right. \cr
 & \left. 
   - \left( \mu_I \leftrightarrow  \nu_I,  W_0 \leftrightarrow W_1 \right)  \right\}^{\prime}, \cr
D & : = \frac{\sqrt{\beta}}{g_s} W_0^{\prime}(z) \frac{1}{\alpha \beta}
\left\{ \sum_I \frac{1}{(z-\mu_I)^2} + 
 2\beta \sum_{I \neq K} \frac{1}{z- \mu_I} \frac{1}{\mu_I -\mu_K}
 + \frac{\sqrt{\beta}}{g_s} \sum_I \frac{W_0^{\prime}(\mu_I)}{z- \mu_I} \right\} \cr
 & -\frac{\sqrt{\beta}}{g_s} \left(\sum_{I, K} 
 \frac{W_0^{\prime}(z) - W_0^{\prime}(\mu_I)}{z - \mu_I} 
 \frac{1}{\mu_I - \nu_K}\right)  \cr
 & - \left( \mu_I \leftrightarrow  \nu_I,  W_0 \leftrightarrow W_1 \right).
\end{split}
\ee

 This is a complicated equation but let us multiply by $g_s^3 \sqrt{\beta}$
  and take the planar limit.

Let
\be
u(z):= \omega_0(z) - \omega_1(z), \qq
v(z):= \omega_0(z) + \omega_1(z).
\ee
 Also, let
 \be
 \begin{split}
  & h_0(z) := 
\lim 
\left\langle\!\!\!\left\langle
 \beta g_s^2 \sum_{I \neq K} 
 \frac{W_0^{\prime}(z) - W_0^{\prime}(\mu_I)}{z - \mu_I} 
 \frac{1}{\mu_I - \mu_K}
\right\rangle\!\!\!\right\rangle,
\cr
  & h_1(z) :=  \lim  
\left\langle\!\!\!\left\langle
\beta  g_s^2 \sum_{I \neq K} 
 \frac{W_1^{\prime}(z) - W_1^{\prime}(\nu_I)}{z - \nu_I} 
 \frac{1}{\nu_I - \nu_K}
\right\rangle\!\!\!\right\rangle,
\cr
  & g_0(z) := \lim 
\left\langle\!\!\!\left\langle
\beta g_s^2 \sum_{I, K} 
 \frac{W_0^{\prime}(z) - W_0^{\prime}(\mu_I)}{z - \mu_I} 
 \frac{1}{\mu_I - \nu_K}
\right\rangle\!\!\!\right\rangle,
\cr
  & g_1(z) := \lim  
\left\langle\!\!\!\left\langle
\beta g_s^2 \sum_{I, K} 
 \frac{W_1^{\prime}(z) - W_1^{\prime}(\nu_K)}{z - \nu_K} 
 \frac{1}{\nu_K - \mu_I}
\right\rangle\!\!\!\right\rangle.
\end{split}
\ee

In terms of $u$, $v$, the planar SD equations can be rewritten as
\bel{uv2}
\begin{split}
&\left( \frac{2+\alpha}{4} \right) u(z)^2 
+\left( \frac{2-\alpha}{4} \right) v(z)^2 \cr
&+ \frac{1}{2} \bigl( W_0'(z) - W_1'(z) \bigr) u(z)
+ \frac{1}{2} \bigl( W_0'(z) + W_1'(z) \bigr) v(z)
- f_0(z) - f_1(z) = 0,
\end{split}
\ee
\bel{uv3}
\begin{split}
& \frac{(3\alpha-2)(\alpha+2)}{12\alpha} u(z)^3
- \frac{(2-\alpha)}{4\alpha} 
\bigl( W_0'(z) - W_1'(z) \bigr) u(z)^2 
- \frac{1}{2\alpha} \bigl\{ (W_0'(z))^2 + (W_1'(z))^2 \bigr\} u(z) \cr
& - \frac{1}{\alpha} 
\bigl\{ (2-\alpha) u(z) + (W_0'(z) - W_1'(z)) \bigr\}
\left( \frac{(2-\alpha)}{4} v(z)^2 + \frac{1}{2} (W_0'(z) + W_1'(z))
v(z) \right) \cr
& + \frac{1}{\alpha} \bigl(W_0'(z) f_0(z) - W_1'(z) f_1(z) \bigr)
- 2 \left( 1 - \frac{1}{\alpha} \right)
(h_0(z) - h_1(z) ) + g_0(z) - g_1(z) = 0.
\end{split}
\ee
Using the planar Virasoro constraint \eqref{uv2},
we can convert \eqref{uv3} into an equation for $u$:
\bel{u3}
\begin{split}
&\frac{(\alpha+2)}{3\alpha} u(z)^3 + \frac{1}{\alpha} 
\bigl( W_0'(z) - W_1'(z) \bigr) u(z)^2
- \frac{1}{\alpha} \bigl\{
W_0'(z) W_1'(z) + (2-\alpha) (f_0(z) + f_1(z)) \bigr\}
u(z) \cr
& + \frac{1}{\alpha} \bigl( W_1'(z) f_0(z) - W_0'(z) f_1(z) \bigr)
- \frac{2(\alpha-1)}{\alpha} 
\bigl( h_0(z) - h_1(z) \bigr) + g_0(z) - g_1(z) = 0.
\end{split}
\ee
For simplicity, we assume $\alpha \neq -2$. 
Let
\be
x(z):=u(z) + \frac{1}{\alpha+2} \bigl( W_0'(z) - W_1'(z) \bigr)
= \omega_0(z) - \omega_1(z) + \frac{1}{\alpha+2} \bigl( W_0'(z) - W_1'(z) \bigr).
\ee
The cubic equation \eqref{u3} becomes
\be
x(z)^3 - p(z) x(z) - q(z) = 0,
\ee
where
\be
p(z) = \frac{3}{(\alpha+2)^2} \bigl\{ (W_0'(z))^2 + (W_1'(z))^2
+ \alpha W_0'(z) W_1'(z) \bigr\}
+ \frac{3(2-\alpha)}{\alpha+2}
\bigl( f_0(z) + f_1(z) \bigr),
\ee
\be
\begin{split}
q(z) &= - \frac{1}{(\alpha+2)^3} \bigl( W_0'(z) - W_1'(z) \bigr)
\bigl\{ 2 (W_0'(z))^2 + 2 (W_1'(z))^2 + (3 \alpha+2) W_0'(z) W_1'(z)
\bigr\} \cr
& - \frac{3}{(\alpha+2)^2}
\bigl\{ ( 2 - \alpha) W_0'(z) + 2 \alpha W_1'(z) \bigr\}
f_0(z)
+ \frac{3}{(\alpha+2)^2}
\bigl\{ 2 \alpha W_0'(z) + ( 2 - \alpha) W_1'(z) \bigr\}
f_1(z) \cr
& + \frac{6(\alpha-1)}{\alpha+2}
\bigl( h_0(z) - h_1(z) \bigr)
- \frac{3\alpha}{\alpha+2}
\bigl( g_0(z) - g_1(z) \bigr).
\end{split}
\ee
 
At $\alpha=2$, we get the cubic equation for $A_1^{(1)}$ model:
\bel{cubicA11}
\begin{split}
& x^3 - \frac{3}{16} ( W_0' + W_1')^2 x
+ \frac{1}{32} (W_0'-W_1') \bigl\{ (W_0')^2 + (W_1')^2 + 4 W_0'W_1'
\bigr\} \cr
& \qq + \frac{3}{4} (W_1' f_0 - W_0' f_1) + \frac{3}{2} ( - h_0+h_1+g_0-g_1)
= 0.
\end{split}
\ee
Here
\be
x = \omega_0 - \omega_1 + \frac{1}{4} ( W_0' - W_1').
\ee

At $\alpha=1$, it turns into the loop equation for $A_2$ model
\bel{cubicA2}
\begin{split}
&x^3 - \frac{1}{3} \bigl\{
(W_0')^2 + (W_1')^2 + W_0' W_1' + 3 (f_0+f_1)
\bigr\} x  \cr
& \qq + \frac{1}{27} (W_0'-W_1') \bigl\{
2(W_0')^2 + 2(W_1')^2 + 5 W_0' W_1' \bigr\} \cr
& \qq + \frac{1}{3} (W_0' + 2 W_1') f_0 - \frac{1}{3}(2 W_0' + W_1') f_1 
+ g_0 - g_1 =0,
\end{split}
\ee
Here
\be
x = \omega_0 - \omega_1 + \frac{1}{3} (W_0' - W_1').
\ee
This cubic equation \eqref{cubicA2} can be rewritten as follows:
\be
\bigl(x - t_1(z) \bigr)
\bigl(x - t_2(z) \bigr)
\bigl(x - t_3(z) \bigr)
- f_1(z) \bigl( x - t_3(z) \bigr)
- f_0(z) \bigl( x - t_1(z) \bigr)
- g_1(z) + g_0(z) = 0,
\ee
where
\be
t_1(z) = \frac{1}{3} ( 2 W_1'(z) + W_0'(z) ), \ \ 
t_2(z) = - \frac{1}{3} ( W_1'(z) - W_0'(z) ), \ \
t_3(z) = - \frac{1}{3} ( W_1'(z) + 2 W_0'(z) ).
\ee
This is the form which have been analysed before.

  Finally, let us consider the special case where $\alpha =2$ and
  $W_{0} = -W_{1}$.  In this case,  eq. \eqref{cubicA11} reduces to
\bel{cubicA11special}
 x^3 - \frac{1}{8} (W_0')^3  
- \frac{3}{4} W_0' (f_0 + f_1) - \frac{3}{2} (h_0 - h_1 - g_0 + g_1)
= 0,
\ee
 possessing the symmetry of $x$  rotation by cubic root of unity
 $ x \rightarrow \ex^{\pm \frac{2 \pi i}{3}} x$.
 This drastic simplification is understood as the prescription
 $\sqrt{\beta} \rightarrow -\sqrt{\beta}$ for the second species of eigenvalues 
 $\nu_J,\; (J = 1, 2, \dotsc,  N^{(1)})$.
 Let us introduce  notation
\bel{notationspecialcase}
\begin{split}
  z_I  &= \begin{cases}
\mu_I, & (I =1,2,\dotsc, N^{(0)}), \cr
\nu_{I-N^{(0)}}, & (I = N^{(0)}+1, \dotsc, N^{(0)} + N^{(1)}),
\end{cases} \cr
 {\rm sgn}\, I &= 
\begin{cases}
1, & (I =1,2,\dotsc,  N^{(0)}), \cr
-1, & (I =N^{(0)}+1, \dotsc,  N^{(0)}+N^{(1)}).
\end{cases}
\end{split}
\ee
The partition function in this case can be written as that of the $\beta$ deformation of
 one-matrix model with positive and negative ``charges" in the Coulomb gas analogy:
\be
\begin{split}
Z:= \int \de^{N^{(0)}+N^{(1)}} z 
\prod_{1 \leq I<J \leq N^{(0)}+N^{(1)}} | z_I - z_J |^{2\beta({\rm sgn}\, I)({\rm sgn}\, J)}
\exp\left( \frac{\sqrt{\beta}}{g_s} \sum_{I=1}^{N^{(0)}+N^{(1)}} ({\rm sgn}\, I) W_0(z_I)
 \right).
\end{split}
\ee
 The entire S-D equations can be formulated in terms of a single resolvent
 $\widehat{\omega}(z) := \widehat{\omega}_0(z) -\widehat{\omega}_1(z)$ and two kinds of
  quantum deformations
  $\hat{f}(z):= \hat{f}_0(z) + \hat{f}_1(z)$ and 
  $\hat{h}(z):= (\hat{h}_{0}(z) - \hat{h}_{1}(z)) -  (\hat{g}_{0}(z) - \hat{g}_{1}(z))$,
  all of which are written succinctly in this one-matrix notation.
 
\section{ $A_{n}^{(1)}$ affine quiver matrix model}

The partition function for the $\beta$-deformed $A^{(1)}_n$ quiver matrix model is 
defined by
\be
Z:= \int d \lambda \, \ex^{-S_{\mathrm{eff}}},
\ee
where
\be
d \lambda= \prod_{i=0}^n \prod_{I=1}^{N^{(i)}}
d \lambda^{(i)}_I,
\ee
\be
\begin{split}
\ex^{-S_{\mathrm{eff}}}
&:= \prod_{i=0}^n \prod_{1 \leq I < J \leq N^{(i)}}
|\lambda^{(i)}_I - \lambda^{(i)}_J |^{2\beta}
\prod_{i=0}^{n} \prod_{I=1}^{N^{(i)}} \prod_{J=1}^{N^{(i+1)}}
|\lambda^{(i)}_I - \lambda^{(i+1)}_J |^{-\beta}  \cr
& \qq \times 
\exp\left( \frac{\sqrt{\beta}}{g_s} \sum_{i=0}^n 
\sum_{I=1}^{N^{(i)}} W_i(\lambda^{(i)}_I) \right),
\end{split}
\ee
with the periodicity of the index $i$:
$\lambda^{(n+1)}_I = \lambda^{(0)}_I$ and
$N^{(n+1)} = N^{(0)}$.
In the following part, we assume this kind of periodicity for the index $i$: 
$i=k+n+1 \equiv k$.

For later convenience, we define the following functions:
\be
\begin{split}
\widehat{\omega}_i(z)&:=\sqrt{\beta} g_s \sum_{I=1}^{N^{(i)}} 
\frac{1}{z - \lambda_I^{(i)}}, \cr
\widehat{R}^{(i)}_{j_1 j_2 \dotsm j_k}(z)&:=
\sqrt{\beta} g_s \sum_{I=1}^{N^{(i)}}
\frac{\xi_{j_1 j_2 \dotsm j_k}^{(i)}(\lambda_I^{(i)})}
{z - \lambda_I^{(i)}}, \cr
\widehat{F}^{(i)}_{j_1 j_2 \dotsm j_k}(z)
&:= \sqrt{\beta} g_s \sum_{I=1}^{N^{(i)}}
\frac{(W_i'(z) - W_i'(\lambda_I^{(i)})) 
\xi^{(i)}_{j_1 j_2 \dotsm j_k} ( \lambda_I^{(i)})}
{z - \lambda_I^{(i)}}, \cr
\widehat{U}^{(i)}_{j_1 j_2 \dotsm j_k}(z)&:=
\sum_{I=1}^{N^{(i)}} 
\frac{\sqrt{\beta} g_s}{z - \lambda_I^{(i)}}
\frac{\partial \xi^{(i)}_{j_1 j_2 \dotsm j_k}(\lambda_I^{(i)})}
{\partial \lambda^{(i)}_I}.
\end{split}
\ee
Here $i,j_1, \dotsc, j_k = 0,1,\dotsc, n$ and
\be
\xi^{(i)}_{j_1 j_2 \dotsm j_k}(\lambda_I^{(i)})
:= \begin{cases}
 \xi^{(i)}_{j_1}(\lambda_I^{(i)})
\xi_{j_2}^{(i)}(\lambda_I^{(i)})
\dotsm \xi_{j_k}^{(i)}( \lambda_I^{(i)}), & ( k \geq 1), \cr
1 & (k=0),
\end{cases}
\ee
with
\be
\xi^{(i)}_j(\lambda_I^{(i)})
:= \begin{cases}
{\displaystyle \sqrt{\beta} g_s 
\sum_{\stackrel{\scr J=1}{(J \neq I)}}^{N^{(i)}}
\frac{1}{\lambda_I^{(i)} - \lambda_J^{(i)}}}, & (j=i), \cr
{\displaystyle \sqrt{\beta} g_s \sum_{J=1}^{N^{(j)}} 
\frac{1}{\lambda_I^{(i)} - \lambda_J^{(j)}}}
= \widehat{\omega}_j(\lambda_I^{(i)}), & (j \neq i).
\end{cases}
\ee
Notice that $\widehat{R}^{(i)}_{j_1 \dotsm j_k}(z)$ with $k=0$ coincide with
$\widehat{\omega}_i(z)$: 
\be
\widehat{R}^{(i)}(z) = \sqrt{\beta} g_s \sum_{I=1}^{N^{(i)}}
\frac{1}{z-\lambda_I^{(i)}} = \widehat{\omega}_i(z), \qq
(k=0),
\ee

Later, we use several identities 
which relate products of $\widehat{\omega}_i(z)$ 
to sums of these functions.
For $\{ j_1, j_2, \dotsm, j_k \}$ all different,
the identity
\be
\prod_{\ell=1}^k \frac{1}{z - \lambda_{I_{\ell}}^{(j_{\ell})} }
= \sum_{\ell=1}^k \frac{1}{z - \lambda_{I_{\ell}}^{(j_{\ell})}}
\prod_{\stackrel{\scr m=1}{( m \neq \ell)}}^{k}
\frac{1}{\lambda^{(j_{\ell})}_{I_{\ell}} - \lambda^{(j_m)}_{I_m}}
\ee
leads to the following identity:
\be
\widehat{\omega}_{j_1}(z) \widehat{\omega}_{j_2}(z)
\dotsm \widehat{\omega}_{j_k}(z)
= \sum_{\ell=1}^{k}
\widehat{R}^{(j_{\ell})}_{j_1 \dotsm j_{\ell-1} j_{\ell+1} \dotsm j_k}(z),
\qq (\{ j_{\ell} \} \ \mbox{all different}).
\ee
If some of indices $j_{\ell}$ coincide, there are $O(g_s)$ corrections:
\bel{omegaid}
\widehat{\omega}_{j_1}(z) \widehat{\omega}_{j_2}(z)
\dotsm \widehat{\omega}_{j_k}(z)
= \sum_{\ell=1}^{k}
\widehat{R}^{(j_{\ell})}_{j_1 \dotsm j_{\ell-1} j_{\ell+1} \dotsm j_k}(z)
+ O(g_s).
\ee
Explicit forms of \eqref{omegaid} for $k=2, 3$ are given by
\bel{wid2}
\widehat{\omega}_i(z) \widehat{\omega}_j(z)
= \widehat{R}^{(i)}_j(z) + \widehat{R}^{(j)}_i(z)
 - \sqrt{\beta} g_s \widehat{\omega}_i'(z) \delta_{ij},
\ee
\be
\begin{split}
\widehat{\omega}_i(z) \widehat{\omega}_j(z) \widehat{\omega}_k(z)
&= \widehat{R}^{(i)}_{jk}(z) + \widehat{R}^{(j)}_{ki}(z)
+ \widehat{R}^{(k)}_{ij}(z), \qq
( i \neq j \neq k \neq i), \cr
\bigl( \widehat{\omega}_i(z) \bigr)^2 \widehat{\omega}_j(z)
&= 2 \widehat{R}^{(i)}_{ij}(z) + \widehat{R}^{(j)}_{ii}(z) 
- \sqrt{\beta} g_s \frac{d}{dz}\left( \widehat{R}^{(i)}_j(z) \right)
+ \sqrt{\beta} g_s \widehat{U}^{(i)}_j(z), \qq
( i \neq j), \cr
\bigl( \widehat{\omega}_i(z) \bigr)^3
&= 3 \widehat{R}^{(i)}_{ii}(z) 
- 3 \sqrt{\beta} g_s \widehat{\omega}_i(z)
\widehat{\omega}_i'(z) - \beta g_s^2 \widehat{\omega}_i''(z) 
+ 3 \sqrt{\beta} g_s 
\widehat{U}^{(i)}_i(z).
\end{split}
\ee

From
\be
\int d \lambda \sum_{I=1}^{N^{(i)}}
\frac{g_s}{\sqrt{\beta}}
\frac{\partial}{\partial \lambda_I^{(i)}}
\left( \frac{ \sqrt{\beta} g_s 
\xi^{(i)}_{j_1 j_2 \dotsm j_k}(\lambda_I^{(i)})}
{z- \lambda_I^{(i)}} \ex^{-S_{\mathrm{eff}}} \right)
=0,
\ee
we find the S-D equations
\bel{LPEQ1}
\begin{split}
& - \frac{g_s}{\sqrt{\beta}} \frac{d}{dz}
\left\langle\!\!\left\langle
\widehat{R}^{(i)}_{j_1 j_2 \dotsm j_k}(z) 
\right\rangle\!\!\right\rangle 
+ \frac{g_s}{\sqrt{\beta}} \left\langle\!\!\left\langle
\widehat{U}^{(i)}_{j_1 j_2 \dotsm j_k}(z)
\right\rangle\!\!\right\rangle \cr
& +2 \left\langle\!\!\left\langle
\widehat{R}^{(i)}_{i \, j_1 j_2 \dotsm j_k}(z)
\right\rangle\!\!\right\rangle
- \left\langle\!\!\left\langle
\widehat{R}^{(i)}_{(i-1) j_1 j_2 \dotsm j_k}(z)
\right\rangle\!\!\right\rangle
- \left\langle\!\!\left\langle
\widehat{R}^{(i)}_{(i+1) j_1 j_2 \dotsm j_k}(z)
\right\rangle\!\!\right\rangle \cr
& + W_i'(z) \left\langle\!\!\left\langle
\widehat{R}^{(i)}_{j_1 j_2 \dotsm j_k}(z) 
\right\rangle\!\!\right\rangle
- \left\langle\!\!\left\langle
\widehat{F}^{(i)}_{j_1 j_2 \dotsm j_k}(z)
\right\rangle\!\!\right\rangle = 0.
\end{split}
\ee

For $k=0$, the S-D equation \eqref{LPEQ1} takes the form
\bel{SD0}
\begin{split}
& - \frac{g_s}{\sqrt{\beta}} \frac{d}{dz}
\left\langle\!\left\langle
\widehat{\omega}_i(z) 
\right\rangle\!\right\rangle  
 +2 \left\langle\!\!\left\langle
\widehat{R}^{(i)}_{i}(z)
\right\rangle\!\!\right\rangle
- \left\langle\!\!\left\langle
\widehat{R}^{(i)}_{i-1}(z)
\right\rangle\!\!\right\rangle
- \left\langle\!\!\left\langle
\widehat{R}^{(i)}_{i+1}(z)
\right\rangle\!\!\right\rangle \cr
& + W_i'(z) \left\langle\!\left\langle
\widehat{\omega}_i(z) 
\right\rangle\!\right\rangle
- \left\langle\!\!\left\langle
\widehat{F}^{(i)}(z)
\right\rangle\!\!\right\rangle = 0.
\end{split}
\ee
Using the identity \eqref{wid2} for $j=i$, we have
\be
2 \widehat{R}^{(i)}_i(z) = \bigl( \widehat{\omega}_i(z) \bigr)^2
+ \sqrt{\beta} g_s \widehat{\omega}_i'(z).
\ee
Substituting this identity into \eqref{SD0}, we have
\bel{LPEQ1a}
\left\langle\!\!\left\langle
\epsilon \widehat{\omega}_i'(z) 
+ \bigl( \widehat{\omega}_i(z) \bigr)^2
- \widehat{R}^{(i)}_{i-1}(z) - \widehat{R}^{(i)}_{i+1}(z)
+ W_i'(z) \widehat{\omega}_i(z) - \widehat{F}^{(i)}(z) 
\right\rangle\!\!\right\rangle = 0,
\ee
where
\be
\epsilon:= \left( \sqrt{\beta} - \frac{1}{\sqrt{\beta}} \right) g_s.
\ee
The identity \eqref{wid2} for $j=i+1$ gives
\be
\widehat{\omega}_i(z) \widehat{\omega}_{i+1}(z)
= \widehat{R}^{(i)}_{i+1}(z) + \widehat{R}^{(i+1)}_{i}(z).
\ee
Summing over $i$, we have
\be
\sum_{i=0}^n \bigl( \widehat{R}^{(i)}_{i+1}(z)
+ \widehat{R}^{(i)}_{i-1}(z) \bigr) = \sum_{i=0}^{n} 
\widehat{\omega}_i(z) \widehat{\omega}_{i+1}(z).
\ee
Only this combination of $\widehat{R}^{(i)}_{i+1}(z)
+ \widehat{R}^{(i)}_{i-1}(z)$ allows an expression in terms of
the resolvents $\widehat{\omega}_j(z)$.
Hence the sum of \eqref{LPEQ1a} over $i$ gives
the ``Virasoro constraint'':
\bel{VC}
\left\langle\!\!\!\left\langle
\sum_{i=0}^n \Bigl( \epsilon \widehat{\omega}_i'(z)
+ \bigl( \widehat{\omega}_i(z) \bigr)^2 - \widehat{\omega}_i(z)
\widehat{\omega}_{i+1}(z) + W_i'(z) \widehat{\omega}_i(z)
- \widehat{F}^{(i)}(z) \Bigr) 
\right\rangle\!\!\!\right\rangle = 0.
\ee

\section{Planar loop equations for $n=2$}

For simplicity, we consider the S-D equations for the $A^{(1)}_n$ model
in the planar limit: $g_s \rightarrow 0$.

Let
\be
R^{(i)}_{j_1 j_2 \dotsm j_k}(z)
:=\lim \Bigl\langle\!\!\Bigl\langle
\widehat{R}^{(i)}_{j_1 j_2 \dotsm j_k}(z) 
\Bigr\rangle\!\!\Bigr\rangle,
\ee
\be
F^{(i)}_{j_1 j_2 \dotsm j_k}(z)
:=\lim \Bigl\langle\!\!\Bigl\langle
\widehat{F}^{(i)}_{j_1 j_2 \dotsm j_k}(z) 
\Bigr\rangle\!\!\Bigr\rangle,
\ee
In the planar limit, the SD equations \eqref{LPEQ1} are given by
\bel{PLPEQ1}
2 R^{(i)}_{i j_1 j_2 \dotsm j_k}(z)
- R^{(i)}_{(i-1) j_1 j_2 \dotsm j_k}(z)
- R^{(i)}_{(i+1) j_1 j_2 \dotsm j_k}(z)
+ W_i'(z) R^{(i)}_{j_1 j_2 \dotsm j_k}(z)
- F^{(i)}_{j_1 j_2 \dotsm j_k}(z) = 0.
\ee

We write explicit constraints for the resolvents (loop equations) in the $A^{(1)}_2$ model.
The planar Virasoro constraint is given by
\bel{Pn2s2}
\begin{split}
& \omega_0^2 + \omega_1^2 + \omega_2^2- \omega_0 \, \omega_1
- \omega_0\, \omega_2 - \omega_1 \, \omega_2  \cr
& + W_0' \, \omega_0 + W_1' \, \omega_1 + W_2' \, \omega_2
- F^{(0)} - F^{(1)} - F^{(2)} = 0.
\end{split}
\ee
The cubic loop equation takes the form
\bel{Pn2s3}
\begin{split}
&\frac{8}{3} ( \omega_0^3 + \omega_1^3 + \omega_2^3)
- \omega_0( \omega_1^2 + \omega_2^2)
- \omega_1 ( \omega_0^2 + \omega_2^2)
- \omega_2( \omega_0^2 + \omega_1^2) - 
2 \omega_0 \, \omega_1\,  \omega_2 \cr
& + W_0'( 3 \omega_0^2 + W_0' \, \omega_0 - F^{(0)})
+ W_1'( 3 \omega_1^2 + W_1' \, \omega_1 - F^{(1)})
+  W_2'( 3 \omega_2^2 + W_2' \, \omega_2 - F^{(2)}) \cr
& - 4 F^{(0)}_0 -  F^{(0)}_1 -  F^{(0)}_2
- 4 F^{(1)}_1 -  F^{(1)}_2 -  F^{(1)}_0
- 4 F^{(2)}_2 -  F^{(2)}_0 -  F^{(2)}_1 =0.
\end{split}
\ee
The quartic loop equation is given by
\bel{Pn2s4}
\begin{split}
& \frac{13}{2} ( \omega_0^4 + \omega_1^4 + \omega_2^4)
- \omega_0 ( \omega_1^3 + \omega_2^3)
- \omega_1 ( \omega_0^3 + \omega_2^3)
- \omega_2 ( \omega_0^3 + \omega_1^3) \cr
& - \frac{3}{2} ( \omega_0^2 \omega_1^2 + \omega_0^2 \omega_2^2
+ \omega_1^2 \omega_2^2) - 3 \omega_0 \omega_1 \omega_2( \omega_1 + \omega_2
+ \omega_2) \cr
& + W_0' \left[ 9 \omega_0^3 + W_0' \left( \frac{9}{2} \omega_0^2
+ W_0' \omega_0 - F^{(0)} \right)
- 7 F^{(0)}_0 - F^{(0)}_1 - F^{(0)}_2 \right] \cr
& + W_1' \left[ 9 \omega_1^3 + W_1' \left( \frac{9}{2} \omega_1^2
+ W_1' \omega_1 - F^{(1)} \right)
- 7 F^{(1)}_1 - F^{(1)}_2 - F^{(1)}_0 \right] \cr
& + W_2' \left[ 9 \omega_2^3 + W_2' \left( \frac{9}{2} \omega_2^2
+ W_2' \omega_2 - F^{(2)} \right)
- 7 F^{(2)}_2 - F^{(2)}_0 - F^{(2)}_1 \right] \cr
& - 13 F^{(0)}_{00} - 5 F^{(0)}_{01} - 5 F^{(0)}_{02}
- F^{(0)}_{11} - 2 F^{(0)}_{12} - F^{(0)}_{22} \cr
& - 13 F^{(1)}_{11} - 5 F^{(1)}_{12} - 5 F^{(1)}_{01}
- F^{(1)}_{22} - 2 F^{(1)}_{02} - F^{(1)}_{00} \cr
& - 13 F^{(2)}_{22} - 5 F^{(2)}_{02} - 5 F^{(2)}_{12}
- F^{(2)}_{00} - 2 F^{(2)}_{01} - F^{(2)}_{11}  = 0.
\end{split}
\ee
The derivation of these constraints is given in Appendix \ref{appB}.

\section*{Acknowledgements}
We thank Nobuhiro Yonezawa for interesting discussion.
The research of H.~I.~ and T.~O.~
is supported in part by the Grant-in-Aid for Scientific Research (2054278, 23540316)
from the Ministry of Education, Science and Culture, Japan.

\appendix

\section{Derivations of  \eqref{ABCDanswer}, \eqref{ABCDdef} }

In this appendix, we outline the derivation of \eqref{ABCDanswer} and \eqref{ABCDdef},
starting from  the second set of S-D equations  eq. \eqref{higherSD1}, 
eq. \eqref{higherSD2} which are  constraints higher than Virasoro constraints.
Eq.  \eqref{higherSD1} reads
\bel{A:higherread}
\begin{split}
& \left\langle\!\!\!\left\langle \sum_{I, K} \frac{1}{(z-\mu_I)^2} \frac{1}{\mu_I -\nu_K}
\right\rangle\!\!\!\right\rangle
 -   \left\langle\!\!\!\left\langle \sum_{I, K} \frac{1}{(z-\mu_I)} \frac{1}{(\mu_I -\nu_K)^2}
\right\rangle\!\!\!\right\rangle  \cr
&+ 2 \beta \left\langle\!\!\!\left\langle
\sum_{I, K} \frac{1}{z - \mu_I} \frac{1}{\mu_I - \nu_K}   \sum_{J(\neq I)} \frac{1}{\mu_I - \mu_J} 
\right\rangle\!\!\!\right\rangle 
 - \alpha \beta 
\left\langle\!\!\!\left\langle
\sum_{I, K, J} 
\frac{1}{z-\mu_I} \frac{1}{\mu_I - \nu_K} \frac{1}{\mu_I - \nu_J}
\right\rangle\!\!\!\right\rangle \cr
& + \frac{\sqrt{\beta}}{g_s} 
\left\langle\!\!\!\left\langle
\sum_{I, K} \frac{W_0'(\mu_I)}{z-\mu_I} \frac{1}{\mu_I - \nu_K}
\right\rangle\!\!\!\right\rangle = 0.
\end{split}
\ee
The counterpart read from eq. \eqref{higherSD2} is given by replacing
$\mu_I$ by $\nu_K$ in eq. \eqref{A:higherread} and we will not spell it out.
Let us subtract this one  from eq. \eqref{A:higherread}, which we refer to as 
eq. $\eqref{A:higherread}_{as}$,
and analyse this in what follows.
We will make a frequent use of the partial fraction formula
\bel{B:pfrac}
  \sum_{i = 1}^{n} \prod_{j (\neq i)}^n  \frac{1}{z_i -z_j} = 0.
\ee
for  a set of  $n$ complex numbers $(z_1, \cdots z_n)$.
For  the developments of this formula  in the context of, see \cite{Imatsuowinf}.
Using  \eqref{B:pfrac} for $(z, \mu_I, \nu_K, \nu_J)$ and  for $(z, \nu_K, \mu_I, \mu_J)$, 
we convert the fourth term of eq. $\eqref{A:higherread}_{as}$ into
\bel{E:fourth}
\begin{split}
 & 2\alpha \beta \cdot
 \left\langle\!\!\!\left\langle
\sum_{\stackrel{\scr I, K, J}{( K \neq J )}} 
\frac{1}{z-\nu_K} \frac{1}{\nu_K - \nu_J} \frac{1}{\nu_K - \mu_I}
  - 
\sum_{\stackrel{\scr I, K, J}{( I \neq J )}} 
\frac{1}{z-\mu_I} \frac{1}{\mu_I - \mu_J} \frac{1}{\mu_I - \nu_K}
\right\rangle\!\!\!\right\rangle \cr
&   -\alpha \beta  \cdot
 \left\langle\!\!\!\left\langle
\sum_{\stackrel{\scr I, K, J}{( K \neq J )}} 
\frac{1}{z-\mu_I} \frac{1}{z - \nu_K} \frac{1}{z - \nu_J}
  -
\sum_{\stackrel{\scr I, K, J}{( I \neq J )}} 
\frac{1}{z-\mu_I} \frac{1}{z - \mu_J} \frac{1}{z - \nu_K}
\right\rangle\!\!\!\right\rangle \cr
& -\alpha \beta  \cdot
 \left\langle\!\!\!\left\langle
\sum_{I, K}  \frac{1}{z-\mu_I} \frac{1}{(\mu_I - \nu_K)^2}
  -
\sum_{I, K} 
\frac{1}{z-\nu_K} \frac{1}{(\nu_K - \mu_I)^2}
\right\rangle\!\!\!\right\rangle.
\end{split}
\ee
  The first line of eq. \eqref{E:fourth} combined with
   the third term in eq. $\eqref{A:higherread}_{as}$ gives
\bel{F:third+firstoffour}
  2(1- \alpha) \beta \cdot
  \left\langle\!\!\!\left\langle
\sum_{\stackrel{\scr I, K, J}{( I \neq J )}} 
\frac{1}{z-\mu_I} \frac{1}{\mu_I - \mu_J} \frac{1}{\mu_I - \nu_K}
  -
\sum_{\stackrel{\scr I, K, J}{( K \neq J )}} 
\frac{1}{z-\nu_K} \frac{1}{\nu_K - \nu_J} \frac{1}{\nu_K - \mu_I}
\right\rangle\!\!\!\right\rangle .
\ee
The third line of  eq. \eqref{E:fourth} combined with the first and the second terms 
in eq. $\eqref{A:higherread}_{as}$ 
   gives
\bel{H:first+second+thirdoffour}
\begin{split}
&-\frac{1+\alpha \beta}{2} \cdot
 \left\langle\!\!\!\left\langle
\sum_{I, K}  \frac{1}{z-\mu_I} \frac{1}{(z - \nu_K)^2}
  - 
\sum_{I, K} 
\frac{1}{z-\nu_K} \frac{1}{(z - \mu_I)^2}
\right\rangle\!\!\!\right\rangle  \cr
&+ \frac{1-\alpha \beta}{2} \cdot
 \left\langle\!\!\!\left\langle
\sum_{I, K}  \frac{1}{(z-\mu_I)^2} \frac{1}{\mu_I - \nu_K}
 -
\sum_{I, K} 
\frac{1}{(z-\nu_K)^2} \frac{1}{\nu_K - \mu_I}
\right\rangle\!\!\!\right\rangle.
\end{split}
\ee
  Here we have used  $\nu_K$ derivative and $\mu_I$ derivative of
 eq.  \eqref{B:pfrac} for $(z, \mu_I, \nu_K)$.
   All in all, we obtain
  \bel{J:allinall}
  \begin{split}
  & 2(1- \alpha) \beta \cdot
  \left\langle\!\!\!\left\langle
\sum_{\stackrel{\scr I, K, J}{( I \neq J )}} 
\frac{1}{z-\mu_I} \frac{1}{\mu_I - \mu_J} \frac{1}{\mu_I - \nu_K}
  -
\sum_{\stackrel{\scr I, K, J}{( K \neq J )}} 
\frac{1}{z-\nu_K} \frac{1}{\nu_K - \nu_J} \frac{1}{\nu_K - \mu_I}
\right\rangle\!\!\!\right\rangle   \cr
  & - \alpha \beta  \cdot
 \left\langle\!\!\!\left\langle
\sum_{I, J, K} 
\frac{1}{z-\mu_I} \frac{1}{z - \nu_K} \frac{1}{z - \nu_J}
    - 
\sum_{I, K, J} 
\frac{1}{z-\mu_I} \frac{1}{z - \mu_J} \frac{1}{z - \nu_K}
\right\rangle\!\!\!\right\rangle   \cr
  &     -\frac{1- \alpha \beta}{2}  \cdot
 \left\langle\!\!\!\left\langle
\sum_{I, K}  \frac{1}{z-\mu_I} \frac{1}{(z - \nu_K)^2}
  -
\sum_{I, K} \frac{1}{(z - \mu_I)^2}
\frac{1}{z-\nu_K} 
\right\rangle\!\!\!\right\rangle  \cr 
& + \frac{1- \alpha \beta}{2}  \cdot
 \left\langle\!\!\!\left\langle
\sum_{I, K}  \frac{1}{(z-\mu_I)^2} \frac{1}{\mu_I - \nu_K}
  -
\sum_{I, K} 
\frac{1}{(z-\nu_K)^2} \frac{1}{\nu_K - \mu_I}
\right\rangle\!\!\!\right\rangle   \cr
 & + \frac{\sqrt{\beta}}{g_s}  W_0'(z) 
\left\langle\!\!\!\left\langle
\sum_{I, K} \frac{1}{z-\mu_I} \frac{1}{\mu_I - \nu_K}
\right\rangle\!\!\!\right\rangle
 - \frac{\sqrt{\beta}}{g_s}  W_1'(z) 
\left\langle\!\!\!\left\langle
\sum_{I, K} \frac{1}{z-\nu_K} \frac{1}{\nu_K - \mu_I}
\right\rangle\!\!\!\right\rangle  \cr
 &- \frac{\sqrt{\beta}}{g_s} 
\left\langle\!\!\!\left\langle
\sum_{I, K} \frac{(W_0'(z) - W_0'(\mu_I))}{z-\mu_I} \frac{1}{\mu_I - \nu_K}
\right\rangle\!\!\!\right\rangle
 +  \frac{\sqrt{\beta}}{g_s} 
\left\langle\!\!\!\left\langle
\sum_{I, K} \frac{(W_1'(z) - W_1'(\nu_K))}{z-\nu_K} \frac{1}{\nu_K - \mu_I}
\right\rangle\!\!\!\right\rangle  \cr
& =  0.  
\end{split}
  \ee
  
In this expression,  all except the first line take forms which are expressible
in terms of the two resolvents $\widehat{\omega}_{0,1}(z)$, their derivatives and 
polynomials in $z$ once
   we invoke the original  Virasoro constraints eq. \eqref{LEQ0}
\bel{L:LEQ0}
\begin{split}
 \alpha \beta 
\left\langle\!\!\!\left\langle
\sum_{I, J}
\frac{1}{z-\mu_I} \frac{1}{\mu_I - \nu_J} 
\right\rangle\!\!\!\right\rangle  &= \cr
 \left\langle\!\!\!\left\langle
\sum_{I} \frac{1}{(z-\mu_I)^2}
\right\rangle\!\!\!\right\rangle
  &+   2 \beta \left\langle\!\!\!\left\langle
 \sum_{\stackrel{\scr I, J}{( J \neq I)}}
\frac{1}{z-\mu_I} \frac{1}{\mu_I - \mu_J} 
\right\rangle\!\!\!\right\rangle 
+ \frac{\sqrt{\beta}}{g_s} 
\left\langle\!\!\!\left\langle
\sum_{I=1} \frac{W_0'(\mu_I)}{z-\mu_I}
\right\rangle\!\!\!\right\rangle.
\end{split}
\ee
and the one eq. \eqref{LEQ1} obtained by $\mu_I \leftrightarrow \nu_K$.

 In order to handle the first line of eq. \eqref{J:allinall}, let us consider another S-D equation:
 \bel{M:anotherSD}
0 = \int \de^{N^{(0)}} \mu \int \de^{N^{(1)}} \nu
\sum_{I}\frac{\partial}{\partial \mu_I}
\left( \frac{1}{z - \mu_I}
\sum_{J (\neq I)}  \frac{1}{\mu_I - \mu_J}
 \ex^{-S_{\mathrm{eff}}} \right).
\ee
Exploiting eq. \eqref{B:pfrac} for $(z,\mu_I, \mu_J)$,
$(z, \mu_I, \mu_J, \mu_K)$ as well as its $z$ derivative? in eq. \eqref{M:anotherSD}, 
    we reexpress the first line of eq. \eqref{J:allinall}, using  
\bel{P:firstofallinall}
 \begin{split}   
 &  \alpha \beta \left\langle\!\!\!\left\langle
\sum_{\stackrel{\scr I, K, J}{( I \neq J )}} 
\frac{1}{z-\mu_I} \frac{1}{\mu_I - \mu_J} \frac{1}{\mu_I - \nu_K}
\right\rangle\!\!\!\right\rangle  = \;\;\;\;
    -\frac{1}{2} \left\langle\!\!\!\left\langle
    \left(\sum_{I \neq J} \frac{1}{z- \mu_I} \frac{1}{z- \mu_J }\right)^{\prime} 
   \right\rangle\!\!\!\right\rangle  \cr
 & + (2\beta -1) \left\langle\!\!\!\left\langle  
 \sum_{I \neq J}\frac{1}{z- \mu_I}\frac{1}{(\mu_I- \mu_J)^2}  \right\rangle\!\!\!\right\rangle
 + \frac{2\beta}{3} \left\langle\!\!\!\left\langle
 \sum_{I \neq J \neq K \neq I} \frac{1}{(z- \mu_I)(z- \mu_J)(z- \mu_K)} 
 \right\rangle\!\!\!\right\rangle   \cr
 &+ \frac{\sqrt{\beta}}{g_s}\frac{W_0^{\prime}(z)}{2}  \left\langle\!\!\!\left\langle
   \left(\sum_{I \neq J} \frac{1}{z- \mu_I} \frac{1}{z- \mu_J }\right)
   \right\rangle\!\!\!\right\rangle
 - \frac{\sqrt{\beta}}{g_s} \left(\sum_{I \neq J} \left\langle\!\!\!\left\langle
 \frac{W_0^{\prime}(z) - W_0^{\prime}(\mu_I)}{z - \mu_I} 
 \frac{1}{\mu_I - \mu_J}\right) \right\rangle\!\!\!\right\rangle 
  \end{split}   
    \ee
    Substituting this into eq. \eqref{J:allinall}, we obtain the expression
      quoted in the text.


\section{Derivations of the constraints \eqref{Pn2s2}, \eqref{Pn2s3}, \eqref{Pn2s4}}

\label{appB}

\subsection{The constraint \eqref{Pn2s2}}

The planar Virasoro constraint \eqref{Pn2s2} can be obtained by taking
the planar limit of the Virasoro constraint \eqref{VC}.
But we rederive it from the planar
S-D equations \eqref{PLPEQ1} because we will need \eqref{n2k0a}, \eqref{n2k0b} and
\eqref{n2k0c} to obtain higher order loop equations \eqref{Pn2s3} and
\eqref{Pn2s4}.

For $n=2$ and $k=0$, \eqref{PLPEQ1} are explicitly given by
\be
\begin{split}
& 2 R^{(0)}_0 - R^{(0)}_2 - R^{(0)}_1 
+ W_0' \, \omega_0 - F^{(0)} = 0, \cr
& 2 R^{(1)}_1 - R^{(1)}_0 - R^{(1)}_2 + W_1' \, \omega_1 - F^{(1)} = 0, \cr
& 2 R^{(2)}_2 - R^{(2)}_1 - R^{(2)}_0 + W_2' \, \omega_2 - F^{(2)} = 0.
\end{split}
\ee
Notice that we have planar identities:
\be
\omega_i(z) \omega_j(z) = R^{(i)}_j(z) + R^{(j)}_i(z).
\ee
In particular,
\be
R^{(0)}_0 = \frac{1}{2} \omega_0^2, \qq
R^{(1)}_1 = \frac{1}{2} \omega_1^2, \qq
R^{(2)}_2 = \frac{1}{2} \omega_2^2.
\ee
Using these relations, we find
\bel{n2k0a}
R^{(0)}_2 + R^{(0)}_1 = \omega_0^2 + W_0' \, \omega_0 - F^{(0)},
\ee
\bel{n2k0b}
R^{(1)}_0 + R^{(1)}_2 = \omega_1^2 + W_1' \, \omega_1 - F^{(1)}, 
\ee
\bel{n2k0c}
R^{(2)}_1 + R^{(2)}_0 = \omega_2^2 + W_2' \, \omega_2 - F^{(2)}.
\ee
By adding these three constraints, we have the planar Virasoro constraint \eqref{Pn2s2}.

\subsection{The cubic loop equation \eqref{Pn2s3}}

The planar S-D equations \eqref{PLPEQ1} for $n=2$ and $k=1$ are given by
\bel{n2k1a}
2 R^{(0)}_{00} - R^{(0)}_{02} - R^{(0)}_{01}
+ W_0'\, R^{(0)}_0 - F^{(0)}_0 = 0,
\ee
\bel{n2k1b}
2 R^{(0)}_{01} - R^{(0)}_{12} - R^{(0)}_{11}
+ W_0' \, R_1^{(0)} - F^{(0)}_1 = 0,
\ee
\bel{n2k1c}
2 R^{(0)}_{02} - R^{(0)}_{22} - R^{(0)}_{12}
+ W_0'\, R^{(0)}_2 - F^{(0)}_2 = 0,
\ee
and similar six equations obtained by cyclic permutations
of the indices $0 \rightarrow 1 \rightarrow 2 \rightarrow 0$
or $0 \rightarrow 2 \rightarrow 1 \rightarrow 0$.

Using
\be
R^{(0)}_{00}(z) = \frac{1}{3} \omega_0(z)^3,
\ee
\eqref{n2k1a} leads to
\bel{Ps3A}
R^{(0)}_{02} + R^{(0)}_{01} = \frac{2}{3} \omega_0^3
+ \frac{1}{2} W_0'\, \omega_0^2 - F^{(0)}_0.
\ee
The sum of \eqref{n2k1b} and \eqref{n2k1c} gives
\bel{Ps3B}
\begin{split}
R^{(0)}_{11} + 2 R^{(0)}_{12} + R^{(0)}_{22}
&= 2 ( R^{(0)}_{02} + R^{(0)}_{01} )
+ W_0' ( R^{(0)}_2 + R^{(0)}_1 ) - F^{(0)}_1 - F^{(0)}_2 \cr
&= \frac{4}{3} \omega_0^3 + W_0'(2 \omega_0^2 + W_0'\, \omega_0 - F^{(0)})
- 2 F^{(0)}_0 - F^{(0)}_1 - F^{(0)}_2.
\end{split}
\ee
From $\{ 2 \times \eqref{Ps3A} + \eqref{Ps3B} \} +$ (cyclic equations),
we obtain the cubic loop equation \eqref{Pn2s3}.

\subsection{The quartic loop equation \eqref{Pn2s4}}

The explicit form of planar SD equations \eqref{PLPEQ1} for $n=2$ and $k=2$
are given by
\bel{n2k2-1}
2 R^{(0)}_{000} - R^{(0)}_{002} - R^{(0)}_{001} + W_0' R^{(0)}_{00}
- F^{(0)}_{00} =0,
\ee
\bel{n2k2-2a}
2 R^{(0)}_{001} - R^{(0)}_{012} - R^{(0)}_{011} + W_0' R^{(0)}_{01}
- F^{(0)}_{01} =0,
\ee
\bel{n2k2-2b}
2 R^{(0)}_{002} - R^{(0)}_{022} - R^{(0)}_{012} + W_0' R^{(0)}_{02}
- F^{(0)}_{02} =0,
\ee
\bel{n2k2-3a}
2 R^{(0)}_{011} - R^{(0)}_{112} - R^{(0)}_{111} + W_0' R^{(0)}_{11}
- F^{(0)}_{11} =0,
\ee
\bel{n2k2-3b}
2 R^{(0)}_{012} - R^{(0)}_{122} - R^{(0)}_{112} + W_0' R^{(0)}_{12}
- F^{(0)}_{12} =0,
\ee
\bel{n2k2-3c}
2 R^{(0)}_{022} - R^{(0)}_{222} - R^{(0)}_{122} + W_0' R^{(0)}_{22}
- F^{(0)}_{22} =0,
\ee
and similar equations.

The constraint \eqref{n2k2-1} can be rewritten as
\bel{n2k2-1r}
R^{(0)}_{001} + R^{(0)}_{002} = \frac{1}{2} \omega_0^4 
+ \frac{1}{3} W'_0 \omega_0^3 - F^{(0)}_{00}.
\ee
The sum of \eqref{n2k2-2a} and \eqref{n2k2-2b} leads to
\bel{n2k2-2r}
R^{(0)}_{011} + 2 R^{(0)}_{012} + R^{(0)}_{022}
= \omega_0^4 + W_0' \left( \frac{4}{3} \omega_0^3
+ \frac{1}{2} W_0' \omega_0^2 - F^{(0)}_0 \right)
- 2 F^{(0)}_{00} - F^{(0)}_{01} - F^{(0)}_{02}.
\ee
By taking the combination \eqref{n2k2-3a} $ + \ 2 \times $\eqref{n2k2-3b} 
$+$ \eqref{n2k2-3c},
we find
\bel{n2k2-3r}
\begin{split}
& R^{(0)}_{111} + 3 R^{(0)}_{112} + 3 R^{(0)}_{122} + R^{(0)}_{222} \cr
&= 2 \omega_0^4 + W_0' \Bigl\{
4 \omega_0^3 + W_0'( 3 \omega_0^2+ W_0' \omega_0 - F^{(0)} )
- 4 F^{(0)}_0 - F^{(0)}_1 - F^{(0)}_2 \Bigr\} \cr
& - 4 F^{(0)}_{00} - 2 F^{(0)}_{01} - 2 F^{(0)}_{02}
- F^{(0)}_{11} - 2 F^{(0)}_{12} - F^{(0)}_{22}.
\end{split}
\ee
From $\{ 3 \times $ ( \eqref{n2k2-1r} $+$ \eqref{n2k2-2r} ) $+$
\eqref{n2k2-3r}$\} +$ (cyclic equations), we find the quartic loop equation \eqref{Pn2s4}.




\end{document}